\documentclass[letterpaper,pra,aps,showpacs,floatfix,twocolumn]{revtex4-1}
\usepackage{graphicx}
\usepackage[ansinew]{inputenc}
\usepackage{array}
\usepackage{color}
\usepackage{amsmath}
\usepackage{amsxtra}
\usepackage{amstext}
\usepackage{amssymb}
\usepackage{latexsym}
\usepackage{dsfont}
\usepackage{verbatim}
\begin{document}

\title{Back-action-free quantum optomechanics with negative-mass Bose-Einstein condensates}
\author{Keye Zhang$^1$, Pierre Meystre$^2$, and Weiping Zhang$^{1}$}
\affiliation{$^1$Quantum Institute for Light and Atoms, Department of Physics, East China Normal University, Shanghai, P.R.China\\
$^2$B2 Institute, Department of Physics and College of Optical Sciences, University of Arizona, Tucson, Arizona 85721, USA}
\begin{abstract}
We propose that the dispersion management of coherent atomic matter waves can be exploited to overcome quantum back-action in condensate-based optomechanical sensors. The effective mass of an atomic Bose-Einstein condensate modulated by an optical lattice can become negative, resulting in a negative-frequency optomechanical oscillator, negative environment temperature, and optomechanical properties opposite to those of a positive-mass system. This enables a quantum-mechanics-free subsystem insulated from quantum back-action.
\end{abstract}
\maketitle

\section{Introduction}
Atomic Bose-Einstein condensates (BECs) present a number of desirable features for precision measurements as well as for a broad spectrum of tests of fundamental physics. These include, for example, thermal-noise-free sensors for atomic clock and interferometry applications~\cite{Dunningham2005} and high-resolution magnetometers~\cite{Vengalattore2007}, tests of the Casimir-Polder force~\cite{Obrecht2007}, the development of quantum simulators for studies of quantum phase transitions~\cite{Greiner2002} and artificial gauge fields~\cite{Spielman2009}, cavity QED experiments~\cite{Brennecke2007},  and studies of decoherence and quantum entanglement in many-body systems~\cite{Esteve2008, Cramer2013}. These applications benefit significantly from the extremely low temperatures, high-order coherence, and bosonic stimulation properties of BECs. However, the quantum nature of the condensates usually results in quantum back-action that randomly disturbs the quantum state to be detected~\cite{Murch2008, Treutlein2007}, resulting, e.g., in the standard quantum limit (SQL) of displacement measurements~ \cite{Braginsky2}.

Recent experiments have also demonstrated that  in BECs quantum back-action can be suppressed using spin squeezing or particle entanglement caused by atom-atom interactions~\cite{Esteve2008, Gross2010}. This approach is inspired by ideas originally developed in the context of gravitational wave detection~ \cite{Braginsky,Braginsky2}, where the injection of squeezed  light fields in the empty input port of the gravitational wave interferometer was proposed to beat the SQL. However, strong degrees of squeezing and the entanglement of large numbers of particles remain challenging due to their increasing sensitivity to decoherence. 

In this paper we show that the dispersion management of the Schr{\"o}dinger field provides a promising alternative to the elimination of quantum back-action effects in BEC-based measurement schemes. When trapped in a weak optical lattice potential, the condensate can be forced into a regime of anomalous dispersion where it acts as a macroscopic quantum object with negative effective mass~\cite{Eiermann2003}. That negative mass can serve as a back-action canceler to a normal, positive mass partner and isolate quantum-mechanics-free subsystems (QMFSs), as discussed in a recent proposal by Tsang and Caves~\cite{Tsang2012}. A similar noise-canceling effect is also expected to be realized by cavity photons with opposite detunings~\cite{Tsang2010} as well as atomic ensembles with opposite spins~\cite{Wasilewski2010}. 

Cavity optomechanical systems based on the collective motion of BECs~\cite{Brennecke2008} and non-degenerate ultracold atomic gases  \cite{Murch2008} have proven to be particularly well suited to demonstrate a number of quantum effects, including the observation of the quantum back-action of position measurements~\cite{Murch2008}, the asymmetry in the power spectrum of displacement noise due to the noncommuting nature of boson creation and annihilation operators~\cite{Brahms2012}, and the optomechanical cooling of a collective motional mode of an atomic ensemble down to the quantum regime \cite{SSmith2011}. These experiments pave the way to promising ultracold-atoms-based quantum metrology schemes, which we use to illustrate the role of the negative effective mass of the condensate in overcoming the quantum back-action.

\section{Back-Action-Free Quantum Optomechanics}
Reference~\cite{Tsang2012} showed that a simple setup to implement a QMFS comprises two harmonic oscillators, $A$ and $B$, of identical frequencies and opposite masses. In the following we assume that they are coupled optomechanically to a common optical field mode $\hat c$ as well as to time-dependent external perturbations $f_A$ and $f_B$ through the interaction Hamiltonian
\begin{equation}
V= \hbar [\Delta_c + G(\hat{q}+\hat{q}')]\hat{c}^{\dagger}\hat{c}+f_A \hat{q} +f_B \hat{q}'.\label{H1}
\end{equation}
Considering then the variables
\begin{eqnarray}
\hat{Q}&=&\hat{q}+\hat{q}'\,\,\,\,\,\,\,\,\,\,\,\,\,\,\,\,\, \,\,\,\,\hat{P}=\frac{1}{2}(\hat{p}+\hat{p}')\nonumber\\
\hat{\Phi}&=&\frac{1}{2}(\hat{q}-\hat{q}')\,\,\,\,\,\,\,\,\,\,\,\hat{\Pi}=\hat{p}-\hat{p}'
\end{eqnarray}
It is easily verified that $[\hat{Q}, \hat{\Pi}]=0 $ and
\begin{equation}
\dot {\hat{Q}} = \frac{\hat{\Pi}}{m},\,\,\,\dot{ \hat{\Pi} }= -m\omega^2 \hat{Q}+f_B-f_A,\,\,\,\dot{\hat{c}}  = i\Delta_{c}\hat{c}-iG\hat{Q}\hat{c}.\label{dQ}
\end{equation}
so that the dynamical pair of observables formed by the collective position $\hat{Q}$ and relative momentum $\hat{\Pi}$ form a QMFS. 

Equations~(\ref{dQ}) describe the motion of a particle driven by the difference in the external perturbations, $f_B-f_A$, resulting in a frequency shift of the cavity field that can be detected by interferometry. However, unlike the general optomechanical case, since the radiation pressure effect is absent in the equation for $\hat{\Pi}$, this measurement does not introduce any back-action and hence is not subject to the SQL. Complementary conclusions hold for the QMFS characterized by the pair operators $\hat{\Phi}$ and $\hat{P}$ for an optomechanical coupling of the form $G(\hat{q}-\hat{q}')$. In that case the frequency shift is proportional to $f_B+f_A$.  

\begin{figure}[ptbh]
\centering
\includegraphics[width=3.5in]{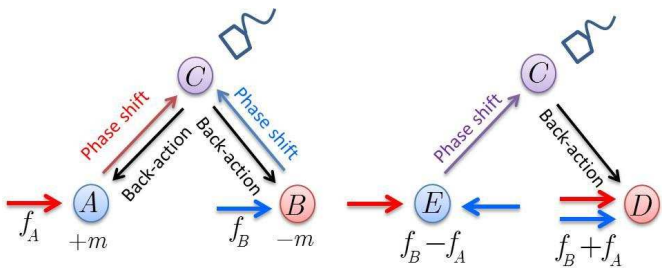}
\caption{(Color online) Relationship diagram for the back-action evading setup in the ``bare'' (left) and ``composite'' (right) representations. The displacement of the composite oscillator $E$ results in a change in the phase of the cavity field $C$ that could be measured by homodyne detection, but the measurement back-action only affects the composite oscillator $D$.   }
\label{loop}
\end{figure}

Further insight into the underlying physics of this back-action-free measurement scheme can be gained by considering the quantum state dynamics. We assume that the system is initially uncorrelated, with the cavity field in a coherent state and the positive-mass oscillator $A$ and negative-mass oscillator $B$ both in their ground state,
\begin{equation}
\left| {\psi (0)} \right\rangle  = {\left| \alpha  \right\rangle _C} \otimes {\left| 0 \right\rangle _A} \otimes {\left| 0 \right\rangle _B}.
\end{equation}
As a result of the optomechanical interaction, (\ref{H1}), the oscillators $A$ and $B$ become entangled with the cavity field $C$. The correlation loop of the total scheme is shown in Fig.\ref{loop}. However, when expressing the state of the system in terms of the composite oscillators $D$ and $E$, described by the operators $\{\hat{Q}, \hat{P}\} $ and $\{\hat{\Phi}, \hat{\Pi}\} $, respectively, we find that it does not suffer three-body entanglement among the subsystems $C$, $D$, and $E$, but only two-body entanglement. Specifically we find, except for an unimportant constant phase factor,
\begin{eqnarray} 
| \psi (t)\rangle  &=& e^{ - |\alpha |^2/2}\sum_n \frac{\alpha^n}{\sqrt {n!}} \exp \left [\frac{-i4nGQ_s}{ \omega }\left (\omega t - \sin \omega t\right )\right ]\nonumber \\
&\times&| n \rangle _C|\phi_n(t) \rangle _D \otimes |\varphi(t)\rangle _E,
\label{state} 
\end{eqnarray}
where
\begin{eqnarray}
\phi_n(t)&=&\frac{-1}{\omega \sqrt{\hbar m \omega}} (f_A+f_B+2\hbar G n)\left ( 1- e^{-i\omega t} \right ),\nonumber \\
\varphi(t)&=&\sqrt{\frac{m\omega}{\hbar}}Q_s\left ( 1- e^{-i\omega t} \right ),\nonumber
\end{eqnarray}
$Q_{s}=(f_B-f_A)/m\omega^2$, and $|n\rangle_C$ are photon Fork states. Equation (\ref{state}) shows that in contrast to the composite oscillator $D$, which becomes entangled with the cavity mode $C$, the composite oscillator $E$ remains uncorrelated with the rest of the system. Rather, it evolves into a time-dependent coherent sate $|\varphi(t)\rangle _E$ that is independent of both the state of the optical field and the composite oscillator $D$. 

The states $|\phi_n(t) \rangle _D$ are $n$-dependent coherent states of the composite oscillator $D$. This is similar to the situation encountered in single-mirror optomechanics~\cite{Knight97}, except for the important dependence of the phase factors on the steady-state displacement $Q_s$ of the oscillator $E$. That dependence makes it easy to read out $Q_s$ without measurement back-action. For example, for $\omega t=2m\pi$, $m$ integer, the state of the system reduces to $| \alpha \exp[-8im\pi G Q_s/\hbar\omega]\rangle _C \otimes | 0 \rangle _D \otimes | 0 \rangle _E$. That is,  the composite oscillators $D$ and $E$ return to the vacuum state---as do the oscillator $A$ and $B$---while the cavity field  becomes a coherent state whose phase could be easily measured by homodyne detection.  

\section{Negative-Mass Oscillators}
We now turn to a discussion of possible realizations of negative-mass optomechanical oscillators.  Negative masses are of course absent in the physical world, but the concept of effective masses---which can, in principle, be negative as well as positive---is familiar from solid-state physics, where it has proven useful in describing the motion of electrons in nonideal lattice potentials~\cite{Callaway1974}. Not surprisingly, this idea has recently been expanded to describe aspects of the quantum wave dynamics of ultracold atoms in optical lattices~\cite{Pu2003}, including Bloch oscillations~\cite{Dahan1996}, the lensing effect in the diffraction of the atomic matter waves~\cite{Eiermann2004}, and the formation of gap solitons~\cite{Fallani2003}. We show in the following how to implement negative effective masses in BEC-based quantum optomechanics  to realize a QMFS that may prove useful in the detection of feeble forces and fields.

Consider for concreteness a scalar atomic BEC confined by both an optical lattice potential $V_{0}\cos^{2}(k_{L}x)$ of periodicity $2\pi/k_L$ and an external trapping potential $U(x)$ that is taken to be slowly varying over the lattice period. Restricting the description to one dimension for simplicity, the condensate is described by the Hamiltonian
\begin{equation}
H=\int\hat{\Psi}^{\dagger}\left[-\frac{\hbar^{2}\nabla^{2}}{2m}+V_{0}\cos^{2}(k_L x)+U(x) \right]\hat{\Psi}dx,\label{eq:Horigin}
\end{equation}
where  $\hat{\Psi}(x)$ is the bosonic field operator of the atomic system and we have neglected inter-atomic collisions. Assuming that the lattice potential is sufficiently shallow that we are far from the Mott insulator transition~\cite{Jaksch1998} we proceed by expanding  $\hat{\Psi}(x)$ in terms of a complete set of basis Bloch functions as
\begin{equation}
\hat{\Psi}(x)=\sum_n\int dq\phi_{n,q}(x)\hat{a}_{n,q}
\end{equation}
where $n$ and $q$ label the band index and the quasimomentum, respectively, and $\hat{a}_{n,q}$ are the associated boson annihilation operators. In the following we assume that the condensate is properly described by the product of an envelope that varies slowly over the period of the lattice, and is characterized by a central wave vector $q_0$, and Bloch functions that capture the rapid oscillations of the condensate caused by optical lattice. We then have approximately~\cite{Pu2003}
\begin{equation}
\phi_{n,q}(x)\approx e^{i(q-q_0)x} \phi_{n, q_0}(x)
\end{equation}
where the mode functions $\phi_{n,q_{0}}(x)$ capture the density oscillations and
\begin{equation}
\hat\Psi(x)=\sqrt{2\pi}\sum_n\phi_{n,q_{0}}(x) \hat{\mathcal A}_{n,q_{0}}(x).
\end{equation}
Here we have introduced the slowly varying bosonic operators (on the scale of the lattice period $2\pi/k_L$)
\begin{equation}
\hat{\mathcal{A}}_{n,q_{0}}(x)=(1/\sqrt{2\pi})\int dqe^{i(q-q_{0})x}\hat{a}_{n,q}
\end{equation}
which describe the dynamics of the condensate envelope in the trapping potential $U(x)$, with 
\begin{equation}
[\hat{\mathcal{A}}_{n,q_{0}}(x),\hat{\mathcal{A}}_{n^{\prime},q_{0}}^{\dagger}(x^{\prime})]=\delta_{n,n^{\prime}}\delta(x-x^{\prime}).
\end{equation}

\begin{figure}
\centering
\includegraphics[width=3in]{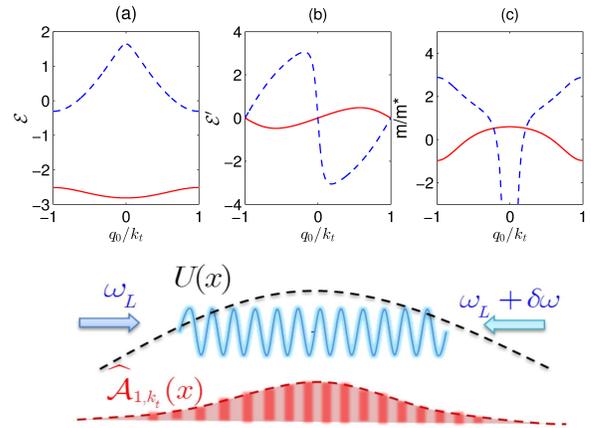}
\caption{(Color online) Top: (a) Bloch energy $\mathcal{E}(q_0)$, (b) its derivative, and (c) the effective mass ratio for a lattice depth $V_0=4.5E_r$ where $E_r=\hbar^{2}k_L^2/2m$ is the atomic recoil energy. Solid (red) lines and dashed (blue) lines are for the first- and second-band case, respectively. Bottom: Sketch of the density profile of a negative-effective-mass condensate in a trap potential U(x): the density is modulated by the optical lattice and peaks at the maximum of U(x).}
\label{band}
\end{figure}

Applying the effective mass method~\cite{Callaway1974} then yields the effective Hamiltonian describing the condensate in the envelope representation as
\begin{eqnarray}
H_{A}&=&\sum_n\int\hat{\mathcal{A}}_{n,q_{0}}^{\dagger}(x)\Big [-\frac{\hbar^{2}\nabla^{2}}{2m_{n,q_{0}}^{*}}+U(x)+\mathcal{E}_{n}(q_{0}) \nonumber \\
&+&  \mathcal{E}_{n}^{\prime}(q_{0})\left(-i\nabla\right) \Big ]\hat{\mathcal{A}}_{n,q_{0}}(x)dx,\label{eq:Ha}
\end{eqnarray}
where the kinetic energy term responsible for the dispersion of the wave packet is modified by the single-particle effective mass
\begin{equation}
m_{n,q_{0}}^*=\hbar^{2}/\mathcal{E}_{n}^{\prime\prime}(q_{0}).
\end{equation}
This corresponds to a parabolic approximation of the energy bands and can be precisely managed by controlling the lattice depth $V_0$, see e.g. Ref.~\cite{Eiermann2003}. Here $\mathcal{E}_{n}^{\prime}(q_{0})$ and  $\mathcal{E}_{n}^{\prime\prime}(q_{0})$ are the first and second-order derivatives of the $n$th-band Bloch energy with respect to the quasimomentum, evaluated at $q_{0}$. Figure~\ref{band} shows that the gradient term  $\mathcal{E}_{n}^{\prime}(q_{0})$ vanishes at the center $(q_{0}=0)$ and edges $(q_{0}=\pm k_{L})$ of the first Brillouin zone. Anomalous dispersion, characterized by a negative effective mass is achieved at the zone edges for odd-$n$ bands, and at the zone center for even-$n$ bands. For deep enough lattices it is sufficient to consider the first band only. This is the situation that we consider in the remainder of this paper. 

The validity of the negative effective mass description relies on the existence of a narrow momentum distribution of the condensate relative to the central wave vector $\pm k_L$. This can be achieved by giving an initial velocity to the condensate, or by considering a condensate initially at rest and adiabatically switching on a moving optical lattice realized by two counter-propagating fields of frequency difference $\delta\omega=2\hbar k_L^{2}/m$. This permits us to neglect the excitation to upper bands by Landau-Zener transitions~\cite{Eiermann2003, Fallani2003}, such that Hamiltonian~(\ref{eq:Ha}) describes a quantum field of negative-effective-mass particles trapped in the potential $U(x)+{\cal E}_1(k_L)$.

The acceleration of particles with a negative mass is opposite the direction of the forces to which they are subjected, so that stability occurs at the maximum of the potential $U(x)$. We thus consider the situation where a condensate of negative effective mass $m_{1, k_L}^*$ is trapped in a potential  $U(x)$ that we approximate as an inverted harmonic potential of frequency  $\Omega=\sqrt{\left|U^{\prime\prime}(x_{1})\mathcal{E}_{1}^{\prime\prime}(k_L)\right|}/\hbar$,  with $x_{1}$ the  position of its maximum. We expand the envelope field operator $\hat{\mathcal{A}}_{1,k_L}(x)$ on the basis of its eigenfunctions $\xi_\ell (x)$, with eigenenergies $\hbar\omega_\ell=-\hbar\Omega(\ell+\frac{1}{2}) < 0$, as 
\begin{equation}
\hat{\mathcal{A}}_{1,k_L}(x)=\sum_\ell \xi_{\ell}(x)\hat{b}_\ell
\end{equation}
where the bosonic operator $\hat{b}_\ell$ annihilates a condensed atom from mode $\ell$.

In contrast with the situation for positive masses, the ``ground mode'' $\xi_{0}(x)$, which has the largest population, now has the highest energy. But since the relative probability of a particle occupying state $\ell$ is proportional to the Boltzmann factor $P_{\ell}\varpropto \exp[-\hbar\omega_{\ell}/k_{B}T]$, we conclude that the effective temperature $T$ of a condensate with negative effective mass is also negative. We note that this situation is closely related to the scheme proposed in Ref.~\cite{Mosk2005} and recently demonstrated by Braun \emph{et al.}~\cite{Braun2013} to achieve negative temperatures. In that case a tight optical lattice was used to establish the density profile of the ultracold atoms initially trapped in a loose harmonic potential $U(x)$. This potential was then reversed from $U(x)$ to $-U(x)$, following which the low-energy states of $U(x)$ corresponded to the high-energy states of $-U(x)$. In our approach, in contrast, $U(x)$ is fixed and we move a shallow optical lattice to change the effective mass of the condensate from positive to negative.

\section{Optomechanical Setup}

Our realization of a negative mass optomechanical system follows closely the approach pioneered in Ref.~\cite{Brennecke2008}, with an optical cavity field of wave vector $k_c\ll k_L$ perturbing the center-of-mass motion of the condensate, but with important differences that we discuss later. For short enough times the depletion of the ground mode $\xi_{0}(x)$ remains small and we can describe it classically, $\hat{b}_{0}\approx\sqrt{N}$, with $N$ the total atom number. The cavity-condensate then coupling takes the form of a multimode cavity optomechanical interaction,
\begin{equation}
H=\hbar\omega_{c}\hat{c}^{\dagger}\hat{c}+\sum_\ell \left [G_\ell(\hat{b}_\ell+\hat{b}_\ell^{\dagger})\hat{c}^{\dagger}\hat{c}+\hbar\omega_{M,\ell}\hat{b}_\ell^{\dagger}\hat{b}_\ell \right ]
\label{multimode}
\end{equation}
where the effective optomechanical coupling coefficients are
\begin{equation}
G_\ell=\sqrt{N}\mathcal{D}\int\xi_0^{*}(x)\cos^{2}(k_{c}x-\theta)\xi_\ell(x)dx.
\end{equation}
Here $\mathcal{D}$ is the single-photon potential depth of the cavity field with annihilation operator $\hat{c}$ and frequency $\omega_{c}$ (including the frequency shift due to the condensate mean field),  $\omega_{M,\ell}=\omega_{\ell}-\omega_{0} <0$ are the oscillation frequencies of the ``effective condensate mirrors''~\cite{Brennecke2008}, and $\theta$ is a phase that depends on the position of the maximum of $U(x)$ relative to the optical potential of the cavity field. We found numerically that coupling to the first excited mode $\ell = 1$ can be made dominant by an appropriate choice of that phase, the remaining modes acting as a mechanical reservoir for that mode~\cite{Brahms2012}. (Other situations with another single dominant mode $\ell$, or with two or more modes coupled with comparable strengths, can also be arranged.)

For the inverted harmonic trap considered here the bath temperature is negative, as already mentioned. One can thus expect a reversed asymmetric displacement spectrum $S_{x}(-\omega)$ for the negative-frequency oscillator $\hat{b}_{\ell}$ and hence a reversed optical output spectrum $n_{c}(-\omega)$ for the cavity field \cite{Brahms2012}. This means that the optomechanical properties of a negative-effective-mass oscillator are reversed from the usual case. For example optical cooling is realized by a blue-detuned rather than a red-detuned driving field, and stationary bipartite entanglement is optimized in the red-detuned case \cite{Genes2008}, with the situation possibly even more interesting in the multimode case. When mechanical damping is weaker than the cavity decay, these negative-frequency oscillators can serve as a negative temperature bath for a cavity field of positive frequency $\omega_{c}$, so that it will exhibit gain as in usual laser theory---where a negative temperature is provided by the inversion of the active medium~\cite{QN2004}.

\begin{figure}[ptbh]
\centering
\includegraphics[width=3.5in]{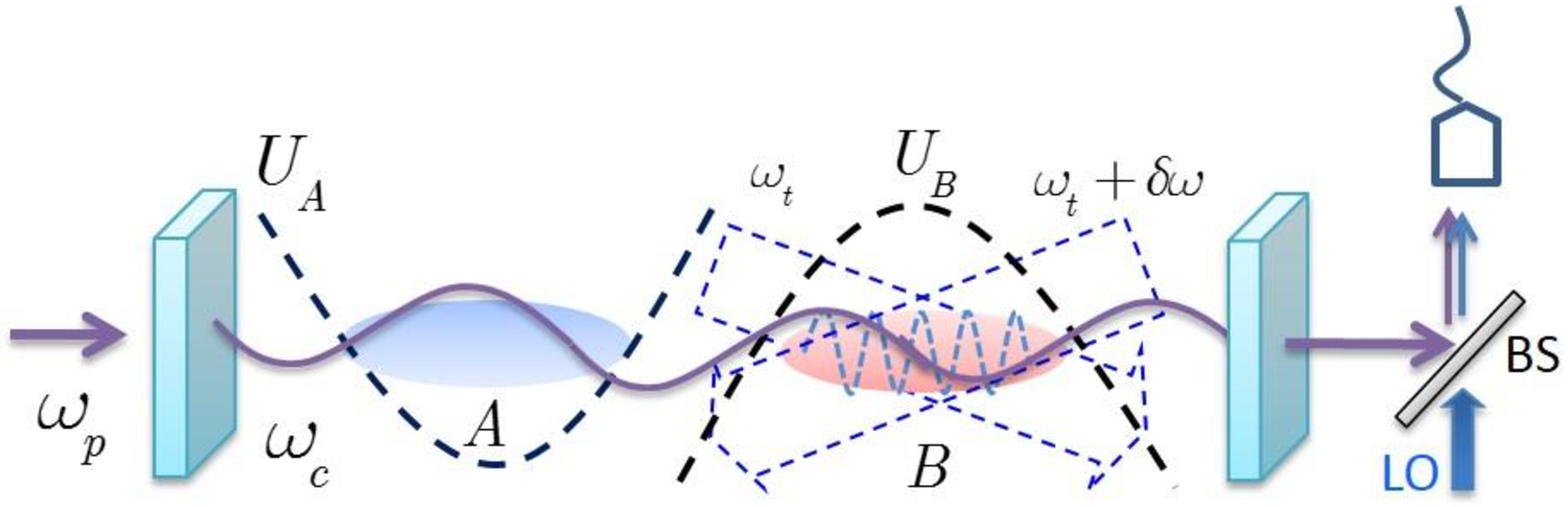}
\caption{(Color online) Possible arrangement for back-action-free field and force detection. Two condensates $A$ and $B$ are trapped along the axis of an optical cavity by the potentials $U_A$ and $U_B$, respectively, with a moving optical lattice that drives condensate $B$ and adjusts its effective mass to a negative value.}
\label{GW}
\end{figure}

Figure \ref{GW} shows a possible back-action-free optomechanical setup based on our considerations. Two cigar-shaped condensates are separately trapped along the $x$ direction, with one of them interacting with a moving optical lattice. The cavity resonance is sensitive to the motion of both condensates, which form an effective ``two-mirror'' cavity optomechanical system. This setup would be suitable to probe a external fields with different strength at positions $A$ and $B$. A back-action-free measurement is realized provided that the trapping potentials and the optical lattice depth are such that the optomechanical parameters of the two condensates are identical except for the sign of their masses. This is the major experimental challenge,  although while a small mismatch will result in imperfect back-action cancellation, it can still improve on the original SQL. Other ways to produce a pair of optomechanical oscillators with opposite effective masses might be using a two-component condensate, with only one component sensitive to the optical lattice, or exciting the condensate to the edges of two adjacent Bloch bands. Two-component condensates allow for the sensing and measurement of fields that couple differently to the two components, such as, e.g., magnetic fields and spinor condensates. 

\section{Conclusion}
Summarizing, we have shown that the anomalous dispersion of BECs in optical lattices permits us to realize situations where quantum back-action is canceled, with potential applications in the measurement of feeble forces and fields. Our discussion centers on the use of collision-less BECs of a spatial extent large compared to the period of the optical lattice. One might ask if it is possible to achieve a similar effect in ultracold, but noncondensed atoms. In that case every atom is localized within a single lattice well, but the phonon mode associated with the center of mass of the sample can still be delocalized and can enter the quantum regime via cavity optomechanical cooling~\cite{SSmith2011}. However, its thermal damping is of the order of $\gamma\sim\omega_{M}$ and is expected to significantly affect the measurement precision of the scheme. In dense condensates, repulsive atom-atom interactions have nontrivial effects and can result, e.g., in the realization of gap solitons, which have been suggested as an attractive potential source for BEC-based quantum metrology \cite{Lee2005}. Future work will discuss in detail the effects of collisions in the proposed system and consider its usefulness in specific applications such as magnetometry and weak force detection. 

\textit{Note added in proof.} Recently, we became aware of a current related work by M. Woolley and A. A. Clerk, arXiv:1304.4059. 

\section{ACKNOWLEDGMENTS}
We thank A. Geraci for insightful comments. This work was supported by the National Basic Research Program of China under Grant No. 2011CB921604, the NSFC under Grant No.~11204084 and No. 11234003, Specialized Research Fund for the Doctoral Program of Higher Education No.~20120076120003, and the Shanghai Natural Science Fund for Youth Scholars under Grant No.~12ZR1443400. P. M. was supported by the DARPA QuASAR and ORCHID programs through grants from the AFOSR and ARO, the U.S. Army Research Office, and the U.S. NSF.


\begin{thebibliography}{10}
\bibitem{Dunningham2005} J. Dunningham, K. Burnett, and W. D. Phillips, Phil. Trans. R. Soc. A {\bf 363}, 2165 (2005); Y. Wang {\it et al.}, Phys. Rev. Lett. {\bf 94}, 090405 (2005).

\bibitem{Vengalattore2007} M. Vengalattore, J. M. Higbie, S. R. Leslie, J. Guzman, L. E. Sadler, and D. M. Stamper-Kurn, Phys. Rev. Lett. {\bf 98}, 200801 (2007).

\bibitem{Obrecht2007} J. M. Obrecht, R. J. Wild, M. Antezza, L. P. Pitaevskii, S. Stringari, and E. A. Cornell, Phys. Rev. Lett. {\bf 98}, 063201 (2007).

\bibitem{Greiner2002} M. Greiner, O. Mandel, T. Esslinger, T. W. H{\"a}nsch, and I. Bloch, Nature {\bf 415}, 39 (2002).

\bibitem{Spielman2009}  Y.-J. Lin, R. L. Compton, K. Jim{\'e}nez-Garcia, J. V. Porti, and I. B. Spielman, Nature {\bf 462}, 628 (2009).

\bibitem{Brennecke2007} F. Brennecke, T. Donner, S. Ritter, T. Bourdel, M. K{\"o}hl, and T. Esslinger, Nature {\bf 450}, 268 (2007).

\bibitem{Esteve2008} J. Est{\'e}ve, C. Gross, A. Weller, S. Giovanazzi, and M. K. Oberthaler, Nature {\bf 455}, 1216 (2008).

\bibitem{Cramer2013} M. Cramer, A. Bernard, N. Fabbri, L. Fallani, C. Fort,	S. Rosi, F. Caruso, M. Inguscio and M.B. Plenio, Nature Communications, \textbf{4}, 2161 (2013).

\bibitem{Murch2008} K. W. Murch, K. L. Moore, S. Gupta, and D. M. Stamper-Kurn, Nature Physics \textbf{4}, 561 (2008).

\bibitem{Treutlein2007} P. Treutlein, D. Hunger, S. Camerer, T. W. H{\"a}nsch, and J. Reichel, Phys. Rev. Lett. {\bf 99}, 140403 (2007).

\bibitem{Braginsky2} V. B. Braginsky and F. Ya. Khalili, {\em Quantum Measurement,} Cambridge University Press, Cambridge, U.K. (1992).

\bibitem{Gross2010} C. Gross, T. Zibold, E. Nicklas, J. Est{\'e}ve, and M. K. Oberthaler, Nature {\bf 464}, 1165 (2010); M. F. Riedel, P. B{\"o}hi, Y. Li, T. W. H{\"a}nsch, A. Sinatra and P. Treutlein, Nature {\bf 464}, 1170 (2010).

\bibitem{Braginsky} V. B. Braginsky, Y. L. Vorontsov and K. S. Thorne, Science {\bf 209}, 547 (1980); C. M. Caves {\it et al.}, Rev. Mod. Phys. {\bf 52}, 341 (1980); P. Meystre and M. O. Scully, eds., {\em Quantum Optics, Experimental Gravitation, and Measurement Theory,} Plenum Press, New York and London (1983).

\bibitem{Eiermann2003} B. Eiermann, P. Treutlein, Th. Anker, M. Albiez, M. Taglieber, K.-P. Marzlin, and M. K. Oberthaler, Phys. Rev. Lett. \textbf{91}, 060402 (2003).

\bibitem{Tsang2012}M. Tsang and C. M. Caves, Phys. Rev. X {\bf 2}, 031016 (2012).

\bibitem{Tsang2010}M. Tsang and C. M. Caves, Phys. Rev. Lett. \textbf{105}, 123601 (2010).

\bibitem{Wasilewski2010}W. Wasilewski, K. Jensen, H. Krauter, J. J. Renema, M. V. Balabas, and E. S. Polzik , Phys. Rev. Lett. \textbf{104}, 133601 (2010).

\bibitem{Brennecke2008} F. Brennecke, S. Ritter, T. Donner, and T. Esslinger, Science \textbf{322}, 235 (2008).

\bibitem{Brahms2012}N. Brahms, T. Botter, S. Schreppler, D. W. C. Brooks, and D. M. Stamper-Kurn, Phys. Rev. Lett. \textbf{108}, 133601 (2012).

\bibitem{SSmith2011}M. H. Schleier-Smith, I. D. Leroux, H. Zhang, M. A. Van Camp, and V. Vuleti\'c, Phys. Rev. Lett. \textbf{107}, 143005 (2011).

\bibitem{Knight97} S. Mancini, V. I. Man'ko, and P. Tombesi, Phys. Rev. A \textbf{55}, 3042 (1997); S. Bose, K. Jacobs, and P. Knight, ibid. \textbf{56}, 4175 (1997).

\bibitem{Callaway1974}J. Callaway, {\em Quantum Theory of the Solid State}, Academic Press (1974). 

\bibitem{Pu2003} M. J. Steel and W. Zhang, e-print cond-mat/9810284; H. Pu, L. O. Baksmaty, W. Zhang, N. P. Bigelow, and P. Meystre, Phys. Rev. A \textbf{67}, 043605 (2003).

\bibitem{Dahan1996}M. B. Dahan, E. Peik, J. Reichel, Y. Castin, and C. Salomon, Phys. Rev. Lett. \textbf{76}, 4508 (1996).

\bibitem{Eiermann2004} B. Eiermann, Th. Anker, M. Albiez, M. Taglieber, P. Treutlein, K.-P. Marzlin, and M. K. Oberthaler, Phys. Rev. Lett. \textbf{92}, 230401 (2004).

\bibitem{Fallani2003}L. Fallani, F. S. Cataliotti, J. Catani, C. Fort, M. Modugno, M. Zawada, and M. Inguscio, Phys. Rev. Lett. \textbf{91}, 240405 (2003).

\bibitem{Jaksch1998} D. Jaksch, C. Bruder, J. Cirac, C. Gardiner, and P. Zoller, Phys. Rev. Lett. \textbf{81}, 3108 (1998).

\bibitem{Mosk2005}A. Mosk, Phys. Rev. Lett. \textbf{95}, 040403 (2005); A. Rapp, S. Mandt, and A. Rosch, Phys. Rev. Lett. \textbf{105}, 220405 (2010).

\bibitem{Braun2013}S. Braun, J. P. Ronzheimer, M. Schreiber, S. S. Hodgman, T. Rom, I. Bloch, and U. Schneider, Science \textbf{339}, 52 (2013).

\bibitem{Genes2008}C. Genes, A. Mari, P. Tombesi, and D. Vitali, Phys. Rev. A \textbf{78}, 032316 (2008).

\bibitem{QN2004}C. Gardiner and P. Zoller, \emph{Quantum Noise}, Springer, 3rd ed. (2004).

\bibitem{Lee2005}R. Lee, E. Ostrovskaya, Y. Kivshar, and Y. Lai, Phys. Rev. A \textbf{72}, 033607 (2005).

\end{thebibliography}
\end{document}